\documentclass[prb,superscriptaddress,twocolumn,showpacs]{revtex4}
\usepackage{epsfig}
\usepackage{amsmath}
\usepackage{graphicx}
\begin{document}
\title{Few-electron eigenstates of concentric double quantum rings}
\author{B. Szafran}
\affiliation{Departement Fysica, Universiteit Antwerpen,
Groenenborgerlaan 171, B-2020 Antwerpen, Belgium}
\affiliation{Faculty of Physics and Applied Computer Science, AGH
University of Science and Technology, al. Mickiewicza 30, 30-059
Krak\'ow, Poland}
\author{F.M. Peeters}\email{Francois.Peeters@ua.ac.be}
\affiliation{Departement Fysica, Universiteit Antwerpen,
Groenenborgerlaan 171, B-2020 Antwerpen, Belgium}

\date{\today}

\begin{abstract}
Few-electron eigenstates confined in coupled concentric double
quantum rings are studied by the exact diagonalization technique.
We show that the magnetic field suppresses the tunnel coupling
between the rings localizing the single-electron states in the
internal ring, and the few-electron states in the external ring.
The magnetic fields inducing the ground-state angular momentum
transitions are determined by the distribution of the electron
charge between the rings. The charge redistribution is translated
into modifications of the fractional Aharonov-Bohm period. We
demonstrate that the electron distribution can be deduced from the
cusp pattern of the chemical potentials governing the
single-electron charging properties of the system. The evolution
of the electron-electron correlations to the high field limit of a
classical Wigner molecule is discussed.
\end{abstract} \pacs{73.21.La} \maketitle
\section{Introduction}

The phase shift of the electron wave function by the vector
potential\cite{AB} results in oscillations of the quantum
transport properties\cite{Webb,Timp,Fuhrer,Tarucha,Pedersen} of
ring-shaped structures.  The conductance\cite{Buttiker} of metal
and semiconductor rings, is periodic in the external magnetic
field with a period determined by the magnetic flux through the
ring. On the other hand, in bound states of closed circular
quantum rings the single-electron spectrum exhibits periodic
ground-state angular momentum transitions with the period of the
flux quantum.\cite{rev} In confined interacting few-electron
systems fractional Aharonov-Bohm (AB) periodicity of the spectrum
was predicted\cite{CP,CEPL} and subsequently observed in
conductance oscillations measured\cite{PRLK} in a transport
spectroscopy experiment. Discussion of the fractional periodicity
in the context of the strength of the electron-electron
interaction was given in Ref. [\cite{Pi}]. The fractional period
for the interacting electron system is also found in realistic
modelling of InGaAs self-assembled quantum rings.\cite{Climente}

Recently, fabrication of self-assembled strain-free double
concentric GaAs/AlGaAs quantum rings was reported.\cite{Nano}
Concentric coupled quantum ring structures can also be produced by
the atomic force microscope tip oxidation
technique.\cite{PRLK,Fuhrer} In this paper we present an exact
diagonalization study of the properties of few-electron states
confined in concentric quantum rings. In the presence of
inter-ring tunnel coupling the electron wave functions undergo
hybridization forming molecular orbitals similarly as in
artificial molecules formed by lateral\cite{l1,l2,lc1,lc2,wens} or
vertical\cite{v2,v4,b1} coupling of quantum dots. The magnetic
field AB period will be significantly different for the internal
and external rings. Therefore, the question arises what will be
the periodicity of the angular momentum transitions for such
hybridized orbitals.

In the two-electron laterally coupled dots the external magnetic
field enhances the localization of the wave functions in each of
the dots.\cite{lc1} Similar is the effect of the electron-electron
interaction favoring charge segregation. On the other hand, in
concentric rings the electron-electron interaction will favor
localization of the electrons in the external ring while the
diamagnetic term of the Hamiltonian will tend to localize the
electrons in the inner ring. We will show that the redistribution
of the electrons between the rings affects the AB period of the
angular momentum transitions, that can be extracted from
conductance measurements\cite{PRLK} on rings connected to
electrodes. Moreover, the angular momentum transitions result in
characteristic cusp patterns of the chemical potential determining
the single-electron charging of the structure. The alignment of
the chemical potentials of the confined electrons with the Fermi
level of the gate electrode can be detected in capacitance
spectroscopy, which was used earlier to study the electronic
structure of self-assembled quantum rings\cite{Lorke} incorporated
in a charge tunable structure.

The present paper extends our previous work on the coupling
between a quantum dot and a quantum ring.\cite{dir} For a single
quantum ring, the envelope of the single-electron ground state
energy depends only on the strength of the confinement in the
radial direction and not on the radius of the ring. For the radial
ring confinement energy $\hbar \omega$, when the radius of the
ring is large as compared to the range of the radial confinement,
the ground-state envelope is approximately given\cite{dir} by
$\sqrt{(\hbar\omega)^2+(\hbar \omega_c)^2}/2$, where $\omega_c$ is
the cyclotron frequency. Therefore, a continuous evolution of the
electron distribution between the two rings should be expected as
a function of the magnetic field in contrast to the rapid
ground-state charge redistributions found previously for a quantum
dot coupled to a surrounding quantum ring.\cite{dir}

A study related to the present one was presented earlier for two
concentric superconducting rings\cite{Ben} in which the coupling
between the rings was mediated by the magnetic self-field of the
separate rings.

The paper is organized as follows. In Section II we present the
model, the results for the single-electron coupling are given in
Section III and for the interacting electron systems in Section
IV. Section V contains the summary and conclusions.

\section{Theory}

We consider a two-dimensional model of circularly symmetric double
concentric rings with confinement potential taken in the form
\begin{equation}
V(\rho)=\frac{m\omega^2}{2}\min\left((\rho-R_1)^2,(\rho-R_2)^2\right),
\end{equation}
where $m$ is the effective electron band mass, $R_1$ and $R_2$
stand for the internal and external ring radii, $\rho$ is the
distance of the origin, and $\omega$ is the harmonic oscillator
frequency for the lateral confinement of the electrons in each of
the rings. Similar models were previously applied for laterally
coupled dots.\cite{lc1,lc2,wens} In our calculations we take the
GaAs value for the mass $m=0.067 m_0$, the dielectric constant
$\epsilon=12.4$ and assume $\hbar \omega=3$ meV. The adopted
oscillator energy corresponds to a length
$l=\sqrt{2\hbar/m\omega}=27.5$ nm which defines the width $d=2l$
of the considered rings.
 The
Hamiltonian of a single electron in a perpendicular magnetic field
($B$), using the symmetric gauge, is
\begin{equation}
h=-\frac{\hbar^2}{2m}\left(\frac{d^2}{d\rho^2}+\frac{1}{\rho}\frac{d}{d\rho}\right)+\frac{\hbar^2L^2}{2m\rho^2}
+\frac{m\omega_c^2\rho^2}{8}-\frac{1}{2}\hbar\omega_cL+V(\rho),
\end{equation}
where $L$ is the angular momentum of the considered state, and
$\omega_c=eB/m$. In the following we refer to the second, third
and fourth term of the Hamiltonian as the centrifugal, diamagnetic
and the orbital Zeeman terms. We neglect the Zeeman interaction of
the electron spin with the magnetic field, which at high fields
polarizes the spins of the confined electrons. The spin Zeeman
interaction is decoupled from the orbital degree of freedom, it
does not influence the tunnel coupling and can be trivially
accounted for as an energy shift linear in $B$.\cite{dir} The
eigenstates of the $N$-electron Hamiltonian
\begin{equation}
H=\sum_{i=1}^N h_i + \sum_{i=1}^N\sum_{j>i}^N
\frac{e^2}{4\pi\epsilon \epsilon_0 r_{ij}} \label{HN}
\end{equation}
are found with a standard\cite{RM,Maarten} exact diagonalization
approach using the single-electron eigenstates of operator (2) to
construct the basis elements in the form of Slater determinants.
We use the numerical method as originally developed to discuss the
coupling between a quantum dot and a quantum ring.\cite{dir} The
single-electron Hamiltonian (2) is diagonalized using a finite
difference scheme and the Coulomb matrix elements are integrated
numerically.

\begin{figure}[htbp]
\hbox{\epsfxsize=75mm
                \epsfbox[4 146 580 687] {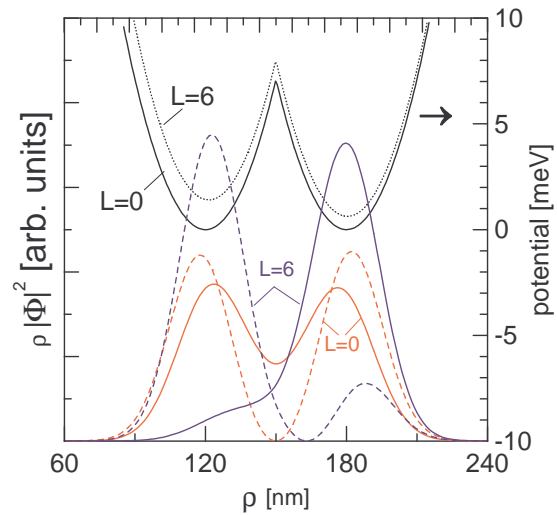}\hfill}
\caption{(color online) Radial profile of the confinement
potential (black solid curve referred to the right vertical axis)
of the two concentric rings for $R_1=120$ nm and $R_2=180$ nm at
$B=0$. The black dotted curve shows the sum of the confinement
potential and the centrifugal potential for $L=6$. Red (light
gray) and blue (dark gray) curves show the square of the modulus
of the two lowest-energy single-electron wave functions multiplied
by Jacobian $\rho$ at $B=0$ for $L=0$ and $L=6$, respectively. The
lower-energy orbitals are given by the solid curves and the
higher-energy orbitals by the dashed curves. } \label{ff}
\end{figure}

\begin{figure}[htbp]
\hbox{\epsfxsize=60mm
                \epsfbox[37 52 570 600] {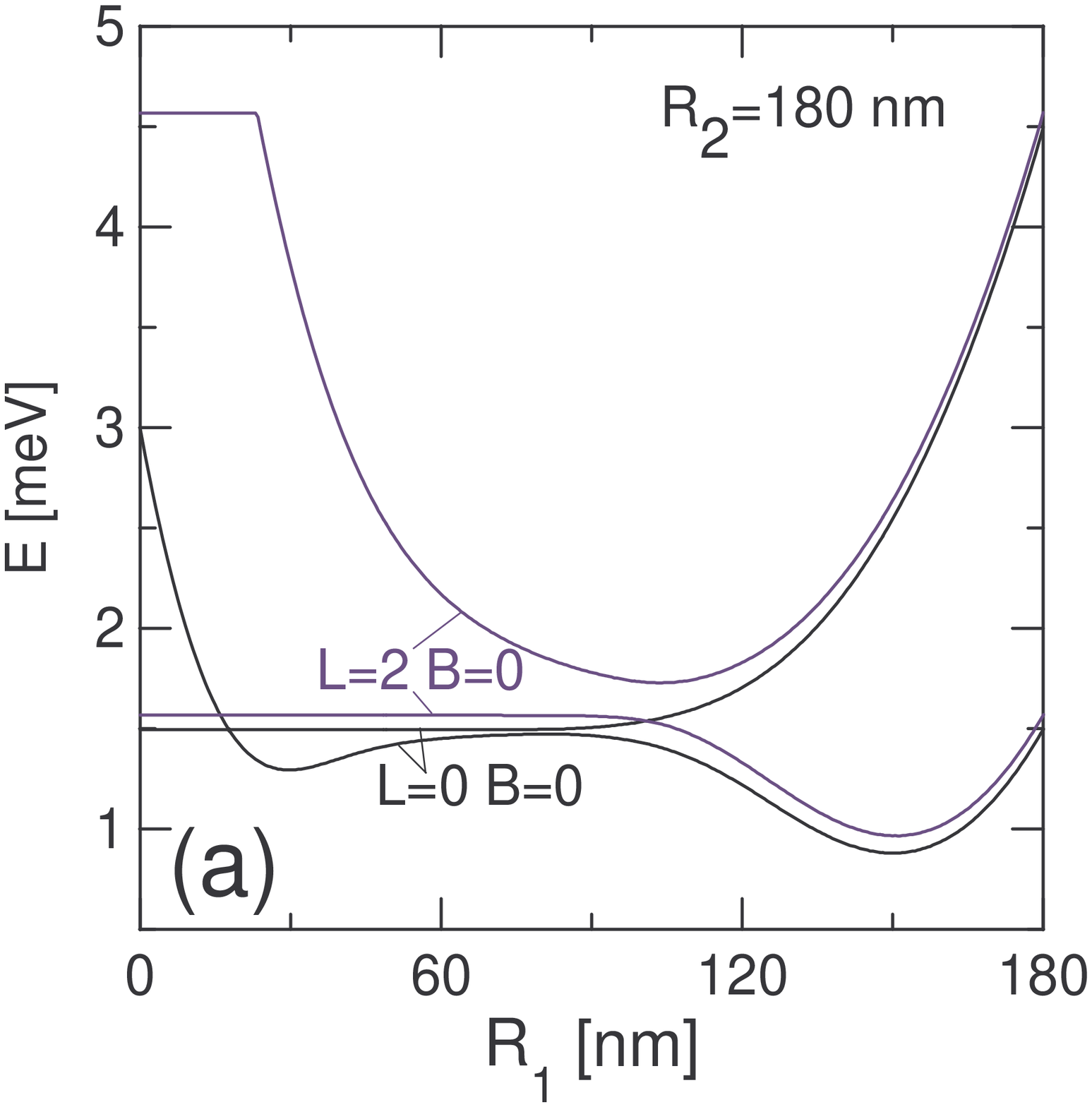}\hfill}
                \hbox{\epsfxsize=60mm
                \epsfbox[37 52 570 600] {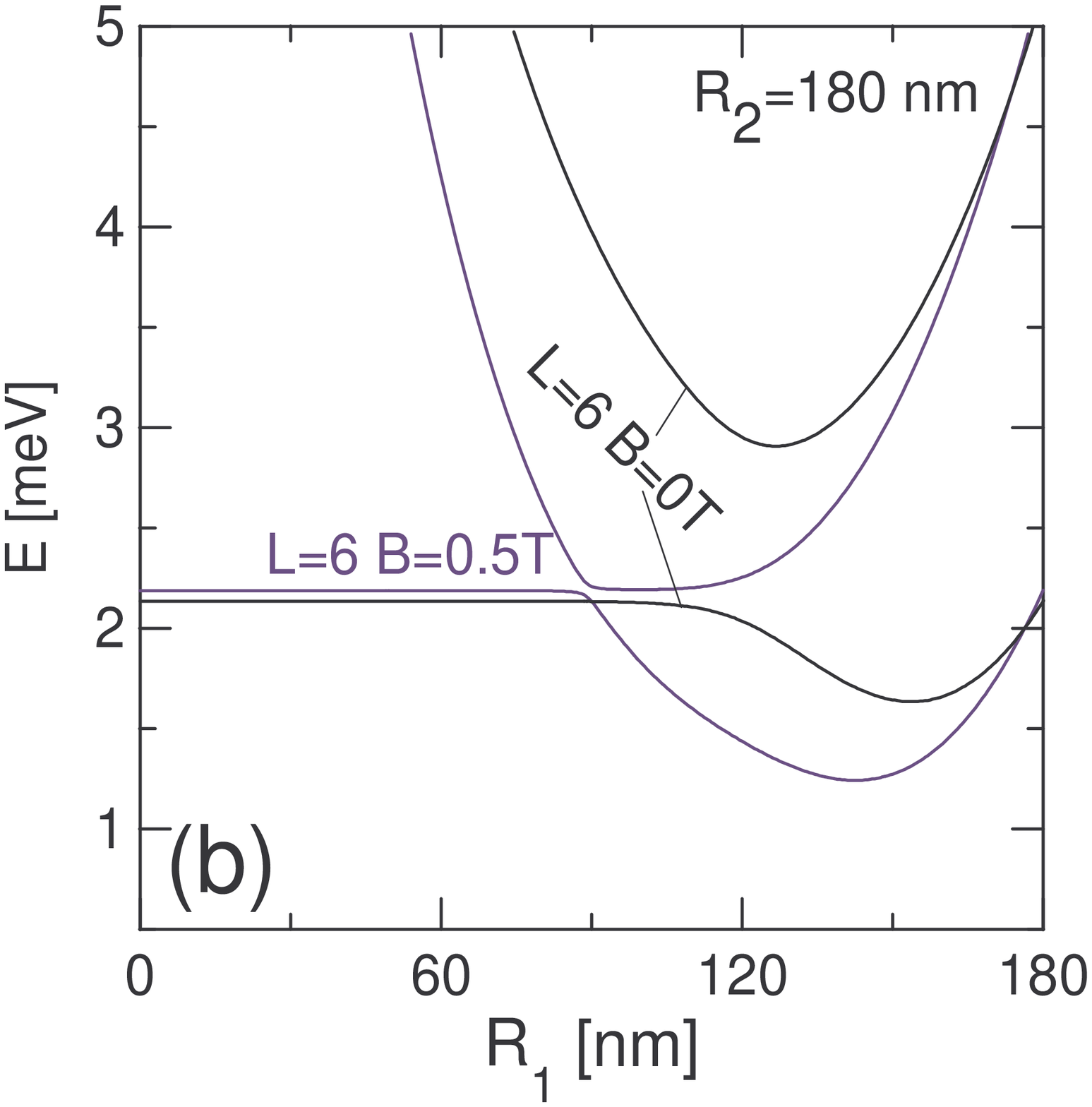}\hfill}
\caption{(color online) Two lowest single-electron energy levels
for $L=0$, and $L=2$ at $B=0$ (a) and for $L=6$ at $B=0$ and 0.5 T
(b), as functions of the internal ring radius for an external ring
of radius $R_2=180$ nm.
 }
\end{figure}

\begin{figure*}[htbp] \hbox{\hbox{\epsfysize=65mm
                \epsfbox[20 68 550 580] {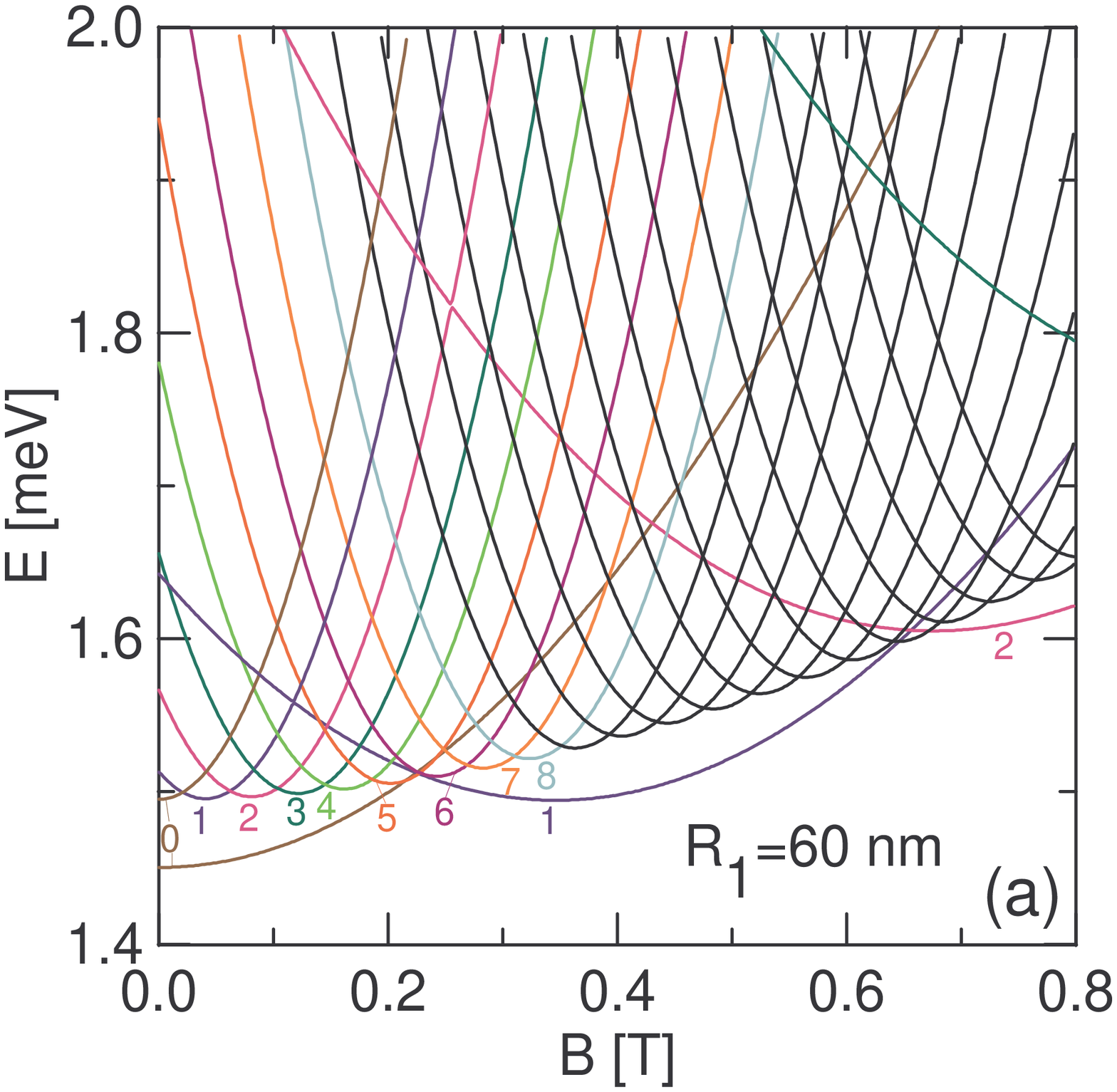}\hfill}
                \hbox{\epsfysize=65mm
                \epsfbox[40 20 570 533] {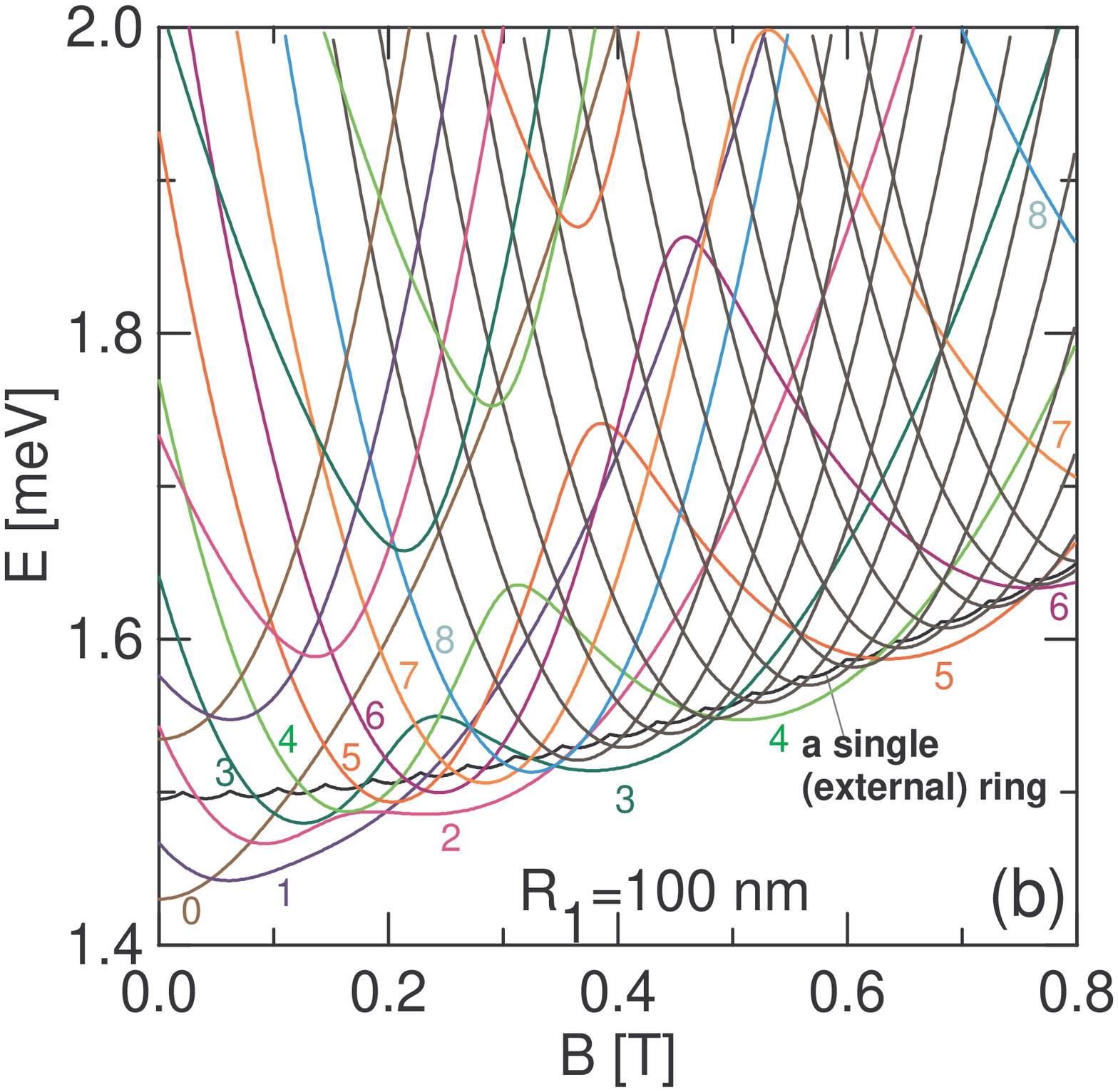}}\hfill}\hbox{
                \hbox{\epsfysize=65mm                 \epsfbox[15 56 545 569] {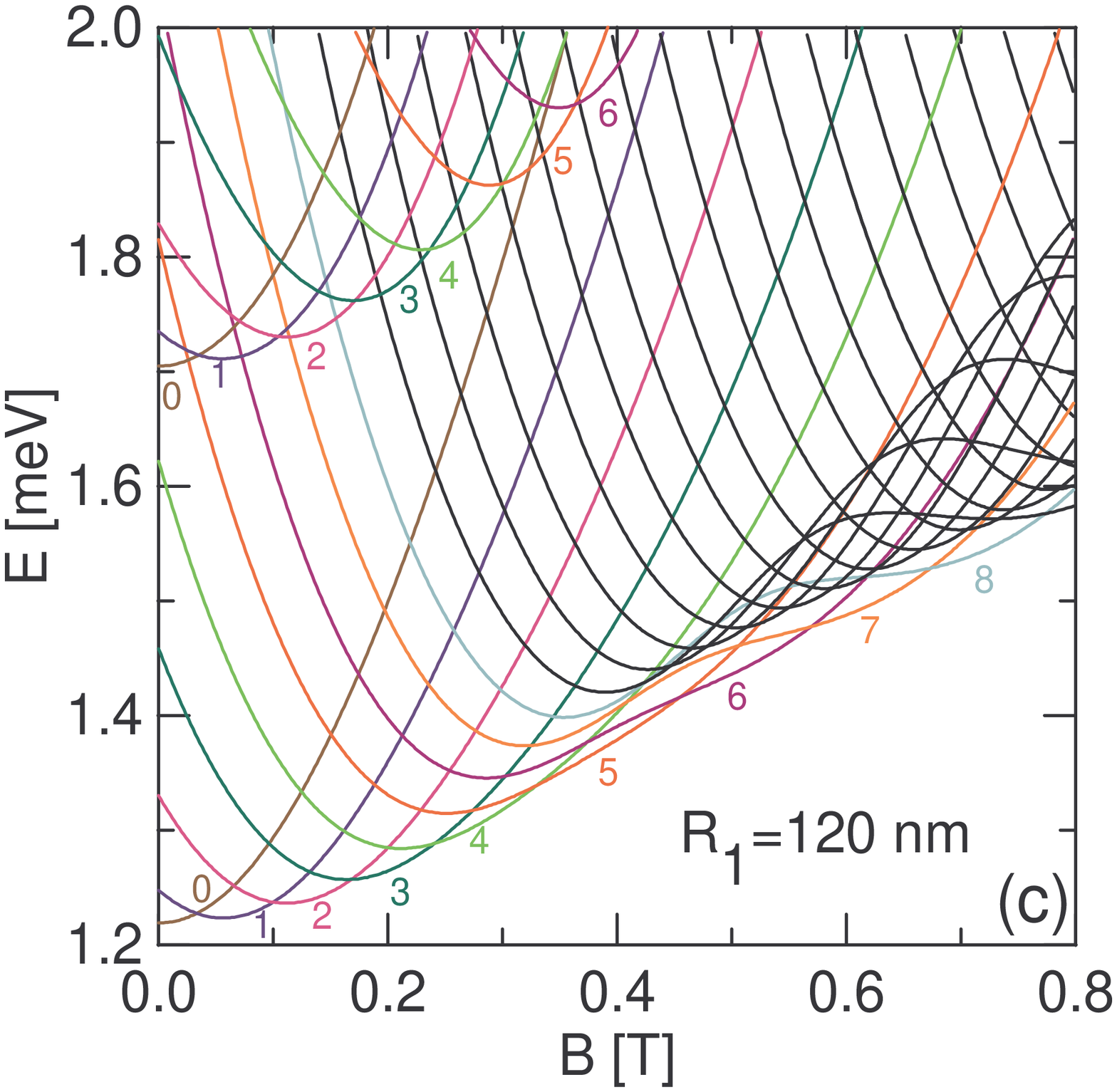}}\hfill
                \hbox{\epsfysize=65mm                 \epsfbox[10 156 550 670] {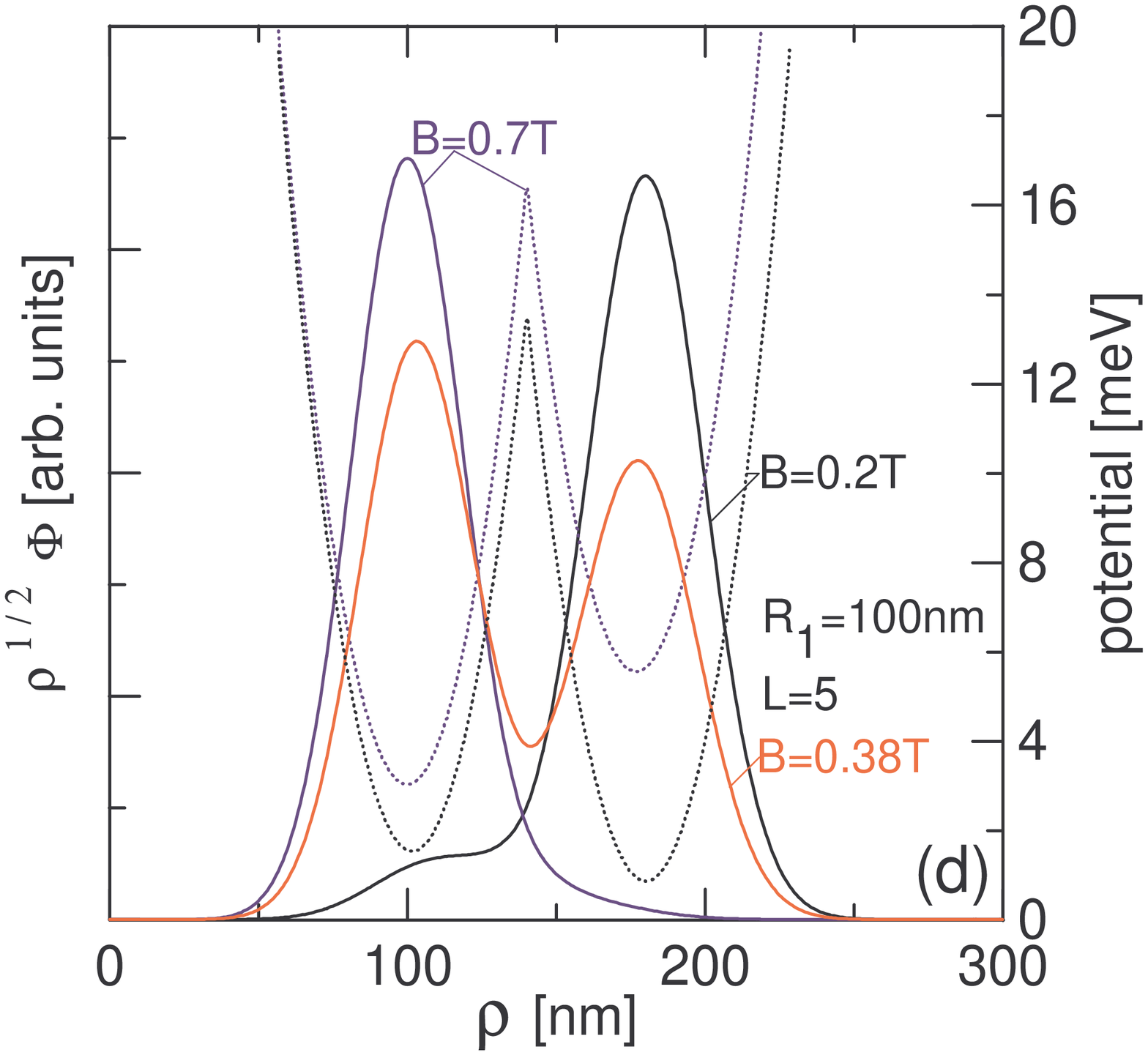}}\hfill
                }
\caption{(color online) (a-c) Single electron spectrum for coupled
rings with the external ring radius $R_2=180$ nm and the internal
ring radius $R_1=60$ (a), 100 (b) and 120 nm (c). Energy levels
corresponding to different angular momenta up to 8 were plotted
with different colors. In (b) the ground state of a single ring
with radius 180 nm is shown by the black curve. (d) Lowest energy
$L=5$ wave function (solid lines) for $R_1=100$ nm, before
($B=0.2$ T) at (0.38 T) and after (0.7 T) the avoided crossings of
the energy levels [cf. red lines in (b)] corresponding to states
localized in the external and internal ring, respectively. Dotted
curves refer to the right vertical axis and show the sum of the
confinement, centrifugal and diamagnetic potentials.}
\end{figure*}

\section{Single electron coupling} Let us first discuss the
single-electron states in the coupled concentric rings.
 Fig. 1 shows the potential felt by an electron in the $L=0$ and
$L=6$ states as well as the lowest-energy orbitals (radial
probability densities) for $R_1=120$ nm and $R_2=180$ nm in the
absence of a magnetic field. In the lowest $L=0$ states the
electron is equally probable to be found in both rings and the
orbitals possess a clear bonding and antibonding character. On the
other hand, for $L\neq 0$, the centrifugal potential pushes the
electrons towards the outer ring. In Fig. 1 we show the result for
$L=6$ which clearly shows that the lowest-energy orbital is
shifted to the external ring. As a consequence, the electron in
the excited-state orbital occupies predominantly the inner ring
and the zero of the wave function is displaced from the center of
the barrier to the external ring. We see that the
bonding-antibonding character of the lowest-energy orbitals
occupying both rings is, for increasing $L$, replaced by a
single-ring type of localization. Therefore, the effect of the
centrifugal potential is to lift the tunnel coupling.

The energy levels are shown in Fig. 2 as functions of the inner
ring radius $R_1$ for fixed $R_2=180$ nm. Note that, for $R_1=0$
the system consists of a quantum dot surrounded by a quantum
ring.\cite{dir} The lowest energy level for $L=0$ and $B=0$ [see
Fig. 2(a)] is then associated with the ring-localized state (of
energy close to $\hbar \omega/2=1.5$ meV) and the excited state
corresponds to an electron confined in the parabolic quantum dot
(of energy $\hbar\omega=3$ meV). For $R_1>0$ the quantum dot is
transformed to a quantum ring. The energy of the orbital, which is
predominantly localized in the inner ring, first goes below
$\hbar\omega/2$ and then returns to this value. Around $R_1=80$ nm
the tunnel coupling appears between the internal and the external
rings leading to an energy gap between the two energy levels.
Finally, for
 a single quantum ring ($R_1=R_2=180$ nm) the spectrum resembles the one-dimensional harmonic
oscillator potential.\cite{dir} For $L=2$ at $R_1=0$ both the
lowest energy levels correspond to orbitals localized in the
external ring. The energies are slightly shifted above $\hbar
\omega/2$ and $3\hbar\omega/2$ by the centrifugal potential. The
internal-ring localized level becomes the first excited state near
$R_1=30$ nm. The centrifugal potential lowers the height of the
inter-ring tunnel barrier (see Fig. 1). Consequently the avoided
crossings between the $L=2$ energy levels ($R_1\simeq 100$ nm) is
visibly larger than for $L=0$. A larger centrifugal shift of the
energy levels and a stronger level interaction, a signature of a
stronger tunnel coupling, is observed for $L=6$ [see Fig. 2(b)].
For $L=6$ and $B=0.5$ T the diamagnetic shift of the
external-ring-confined level is almost exactly cancelled by the
orbital Zeeman term (compare the lowest black and blue curves at
$R_1=0$ in Fig. 2(b)). However, the Zeeman term dominates for the
state localized in the internal ring. As a consequence, the energy
levels change their order in a narrow anticrossing near $R_1=90$
nm.

\begin{figure}[htbp]
\hbox{\epsfxsize=65mm
                \epsfbox[5 140 595 640] {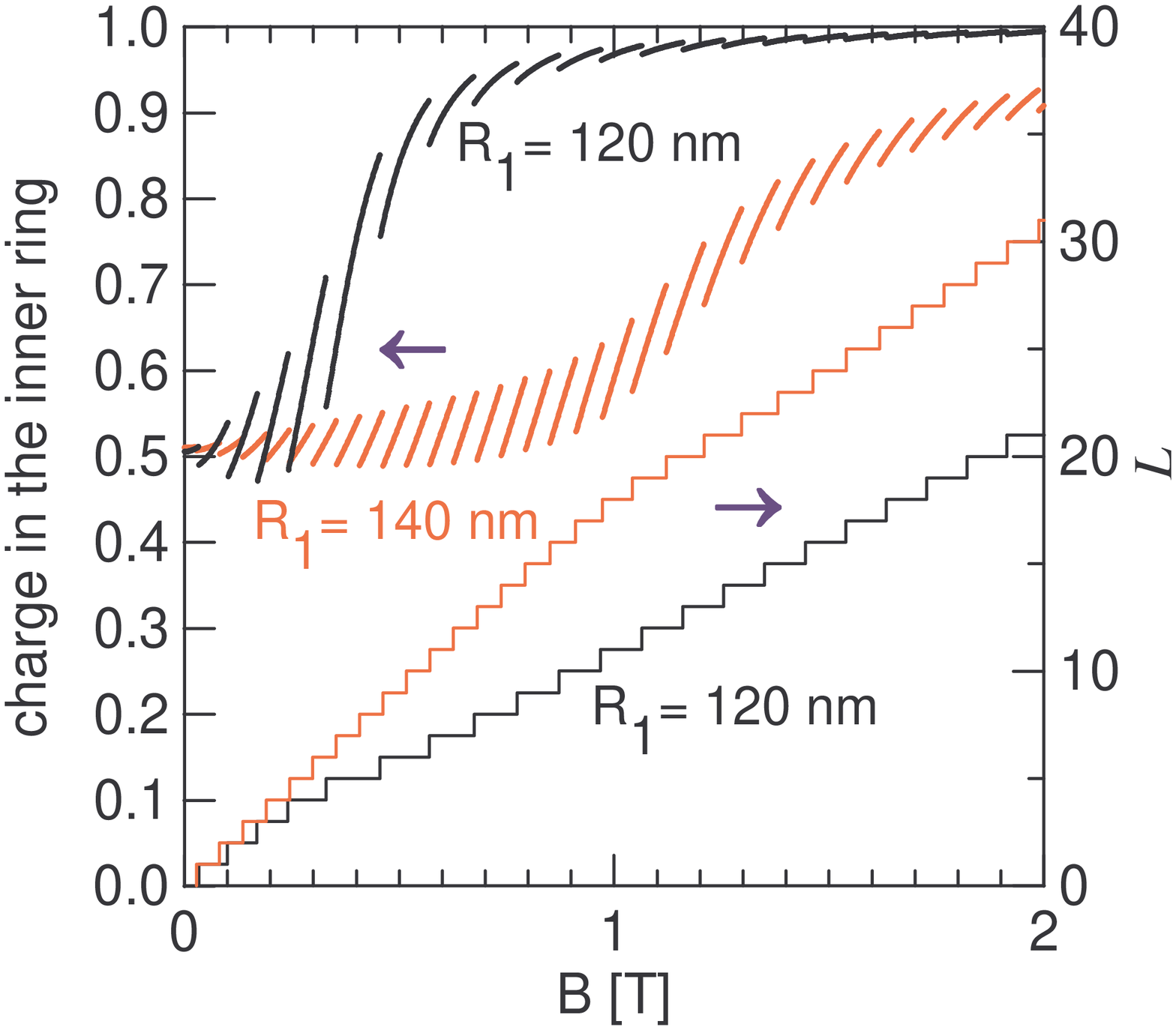}\hfill}
\caption{(color online) The discontinuous lines show the amount of
charge localized in the internal ring for the single-electron
ground-state. The results correspond to the external radius
$R_2=180$ nm and internal radius $R_1=120$ nm (black lines) and
$R_1=140$ nm (red lines) as functions of the magnetic field. The
staircases at the lower part of the figure are referred to the
right axis and show the ground state angular momentum.
 }
\end{figure}

\begin{figure}[htbp] \hbox{\epsfysize=67mm
                \epsfbox[15 31 527 563] {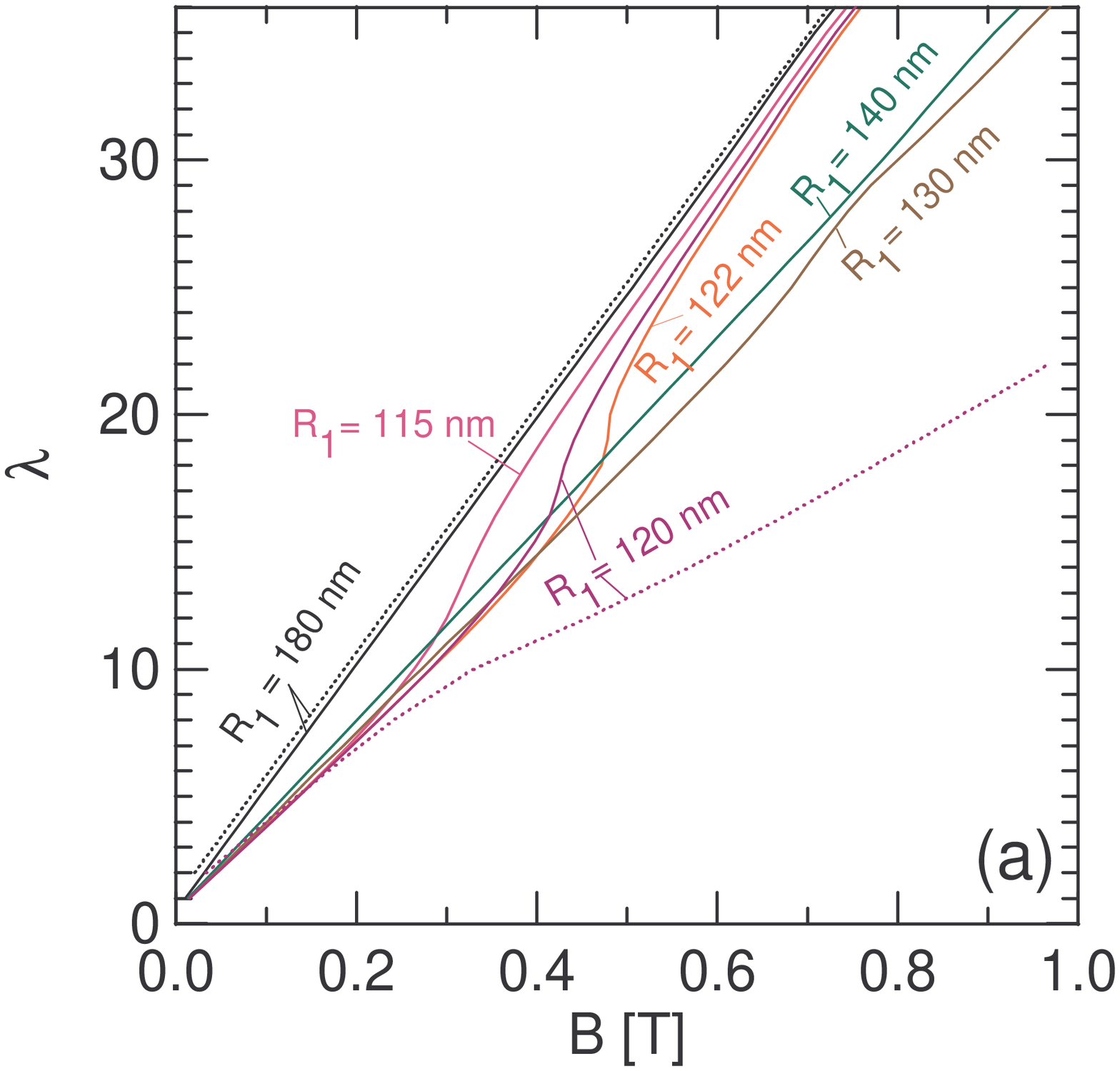}\hfill}
                \hbox{\epsfysize=67mm
                \epsfbox[25 86 546 618] {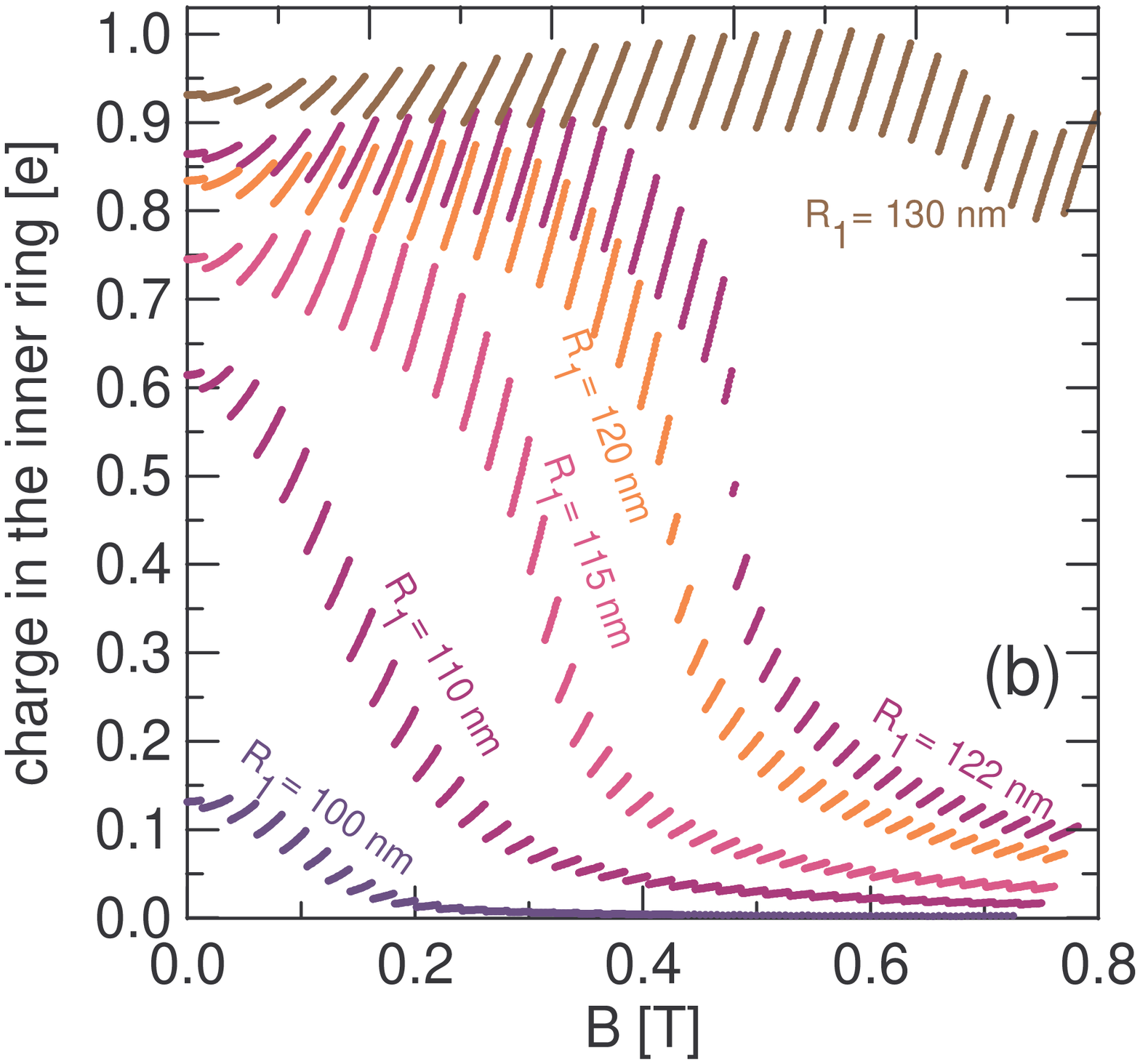}}\hfill
                \hbox{\epsfysize=67mm                 \epsfbox[160 33 678 566] {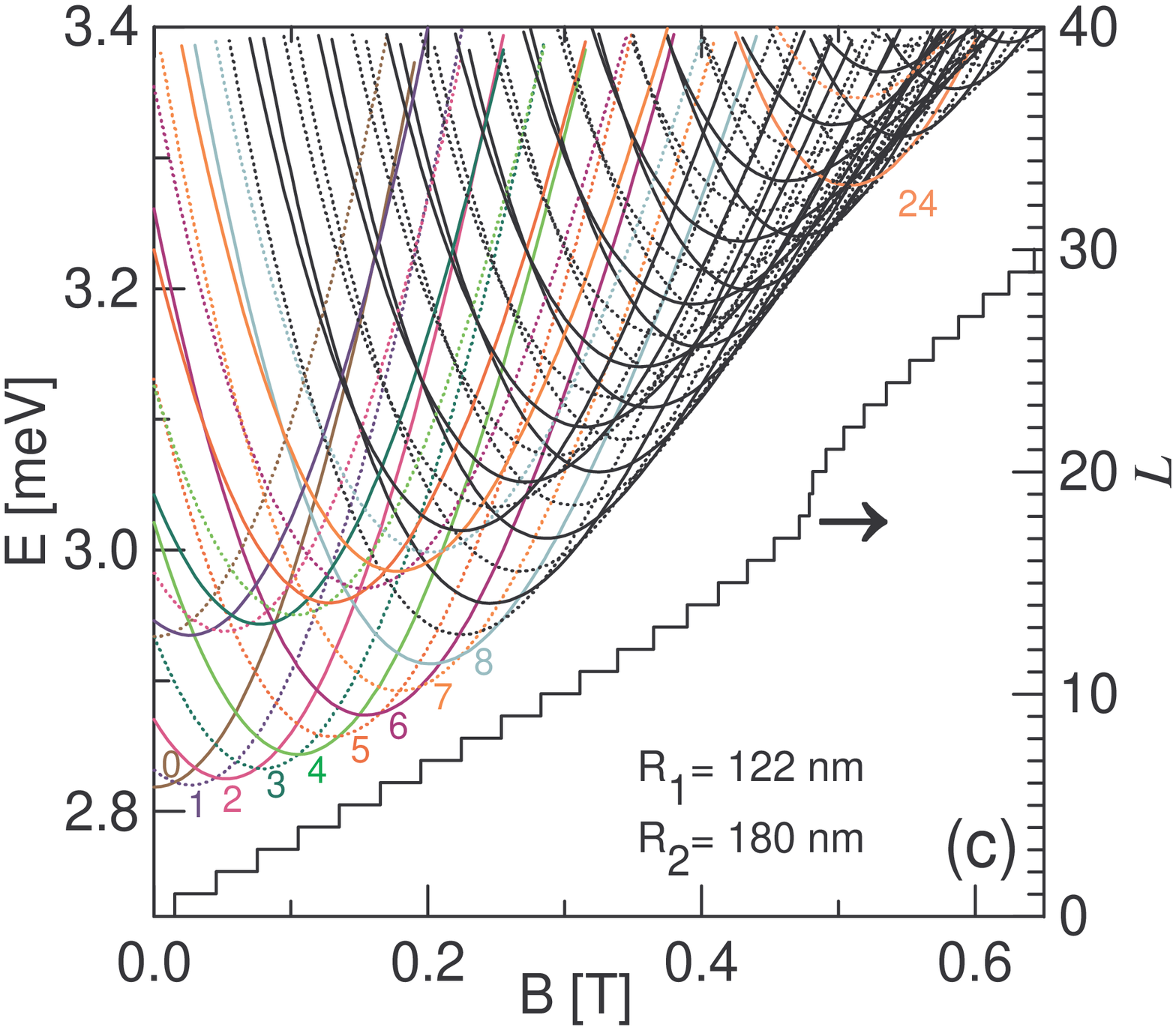}}
\caption{(color online) (a) Upper bound for the two-electron
ground state angular momentum for $R_2=180$ nm and various values
of the inner ring radius. The dotted lines show the values in the
absence of the electron-electron interaction. (b) Charge localized
in the inner ring as function of the magnetic field for $R_2=180$
nm and different radii of the inner ring. (c) The two-electron
energy spectrum for $R_1=122$ nm and $R_2=180$ nm. The spin
singlets are plotted as solid lines and the triplets with dotted
lines. In the bottom of the figure the ground-state angular
momentum staircase is plotted. }
\end{figure}

The dependence of the single-electron energy spectrum on the
external magnetic field is plotted in Figs. 3(a-c) for fixed
$R_2=180$ nm and different internal ring radii.  For $R_1=60$ nm
there is no tunnelling between the rings and the spectrum is a
simple sum of two single ring spectra. The spectrum corresponding
to the internal ring exhibits angular momentum transitions with a
period of $0.214$ T while the period of the one corresponding to
the external ring is $0.0406$ T. These periods correspond to the
flux quantum passing through an effective one-dimensional ring of
radius 55.4 nm and 180 nm, respectively. The ground-state
corresponds to the electron in the internal ring, except for
$B\simeq 0.2$ T and $B\simeq 0.65$ T. The inner-ring localized
states are favored by the $-\frac{1}{\rho}\frac{d}{d\rho}$ term of
the kinetic energy.

For $R_1=100$ nm [see Fig. 3(b)] the inter-ring coupling is
non-negligible. For comparison the ground-state energy of the {\em
single} quantum ring of radius $180$ nm is also shown in Fig. 3(b)
by the black curve. For $B>0.15$ T, sightly above the
ground-state, we observe more frequent angular momentum
transitions than in the ground-state. This energy band corresponds
to the electron predominantly confined in the external ring. With
increasing magnetic field this band closely approaches the single
ring spectrum (cf. the black curve), which indicates that the
electron becomes entirely localized in the external ring. Thus at
high magnetic fields the spectrum of the internal and external
rings become decoupled. Note, that the energy band corresponding
to the localization of the electron in the external ring becomes
distinct only for $L>4$.

Energy levels with the same angular momentum change their order
through avoided crossings. The lowest-energy levels, for $L\geq
2$, possess two minima, after and before the avoided level
crossing. The wave functions and the potentials for the
anticrossing of the $L=5$ energy levels [see the anticrossing of
red lines near 0.38 T at Fig. 3(b)] are presented in Fig. 3(d).
The $L=5$ eigenstate for $B=0.2$ T is the lowest-energy state of
the external ring energy band [see Fig. 3(b) and the paragraph
above] and its wave function is predominantly localized in the
outer ring [see Fig. 3(d)]. At $B=0.38$ T, corresponding to the
smallest distance between the anticrossing energy levels, the
electron can be found with a comparable probability in both rings.
After the avoided crossing the diamagnetic potential localizes the
electron in the internal ring. For $B=0.7$ T the $L=5$ state is
localized almost entirely in the inner ring [see purple curve in
Fig. 3(d)] when it corresponds to the ground-state of the system
[Fig. 3(b)]. Concluding, for $B=0$ and fixed non-zero $L$ the
lowest energy level is predominantly localized in the external
ring due to the centrifugal potential. For high magnetic field the
lowest energy state for a fixed $L$ is transferred to the internal
ring by the diamagnetic term of the Hamiltonian.

For $R_1=120$ nm [Fig. 3(c)] the coupling between the two rings is
stronger and the difference between the centrifugal potentials in
both rings is smaller. Consequently the two decoupled spectra of
the internal and external ring are only distinguishable for
$B>0.5$ T. The amount of electron charge localized in the internal
ring [integrated over $\rho$ from 0 to $(R_1+R_2)/2$] for the
ground-state is plotted in Fig. 4 together with the ground-state
angular momentum. For low magnetic field the ground state wave
functions are almost equally distributed between the two rings and
at high field they are entirely localized in the inner ring.
Consequently, the period of the ground-state oscillations
increases with $B$ (see the slope of the black staircase in Fig.
4). Note, that the decoupling of the spectra in Fig. 3(c) for
$B>0.5$ T ($R_1=120$ nm) is accompanied by the transfer of the
electron to the internal ring (see Fig. 4). For $R_1=140$ nm many
more angular momentum transitions are needed before the electron
becomes entirely localized in the inner ring.

At the end of this Section we would like to explain the role of
the adopted finite value of the rings width for our results. The
studied rings radii ($R\simeq 150$ nm) and width ($d=55$ nm)
correspond to structures produced by the tip oxidation
technique.\cite{Fuhrer} For instance the ring of
Ref.[\cite{Fuhrer}] is characterized by $R=132$ nm and $d=65$ nm.
In the limit of infinite oscillator energy ($\hbar\omega$) the
rings become strictly one-dimensional ($d\rightarrow 0$) and
decoupled due to the infinite interring barrier. The energy levels
of states confined in one-dimensional rings only depend on the
magnetic flux\cite{rev}
$E_i(L)=\frac{\hbar^2}{2mR_i^2}(L-\Phi_i/\Phi_0)^2$, where $i=1,2$
stands for the internal and external ring localization
respectively, $\Phi_0=h/e$ is the flux quantum and $\Phi_i$
corresponds to the flux through the radius $R_i$. It is clear that
the localization of the lowest energy level of a pair of
one-dimensional rings will oscillate abruptly between internal and
external rings when the magnetic field is increased. However, this
switching is deprived of physical consequences since due to the
infinite interring barrier the electron is not allowed to release
its energy tunnelling from one ring to the other. Note, that a
trace of the discussed localization switching can be observed in
Fig. 3(a) for negligible interring tunnel coupling. Decoupled
spectra with short appearances of the external ring localization
in the lowest-energy state similar to Fig. 3(a) are obtained for
$R_1=120$ nm, $R_2=180$ nm for $d$ decreased from 55 nm [as in
Fig. 3(c)] to 13.5 nm ($\hbar \omega=50$ meV). The rapid
localization switching disappears for the nontrivial case of a
non-negligible tunnel coupling [cf. Figs. 3(b-d)].

\begin{figure}[htbp]
\hbox{\epsfxsize=50mm
                \epsfbox[46 90 567 628] {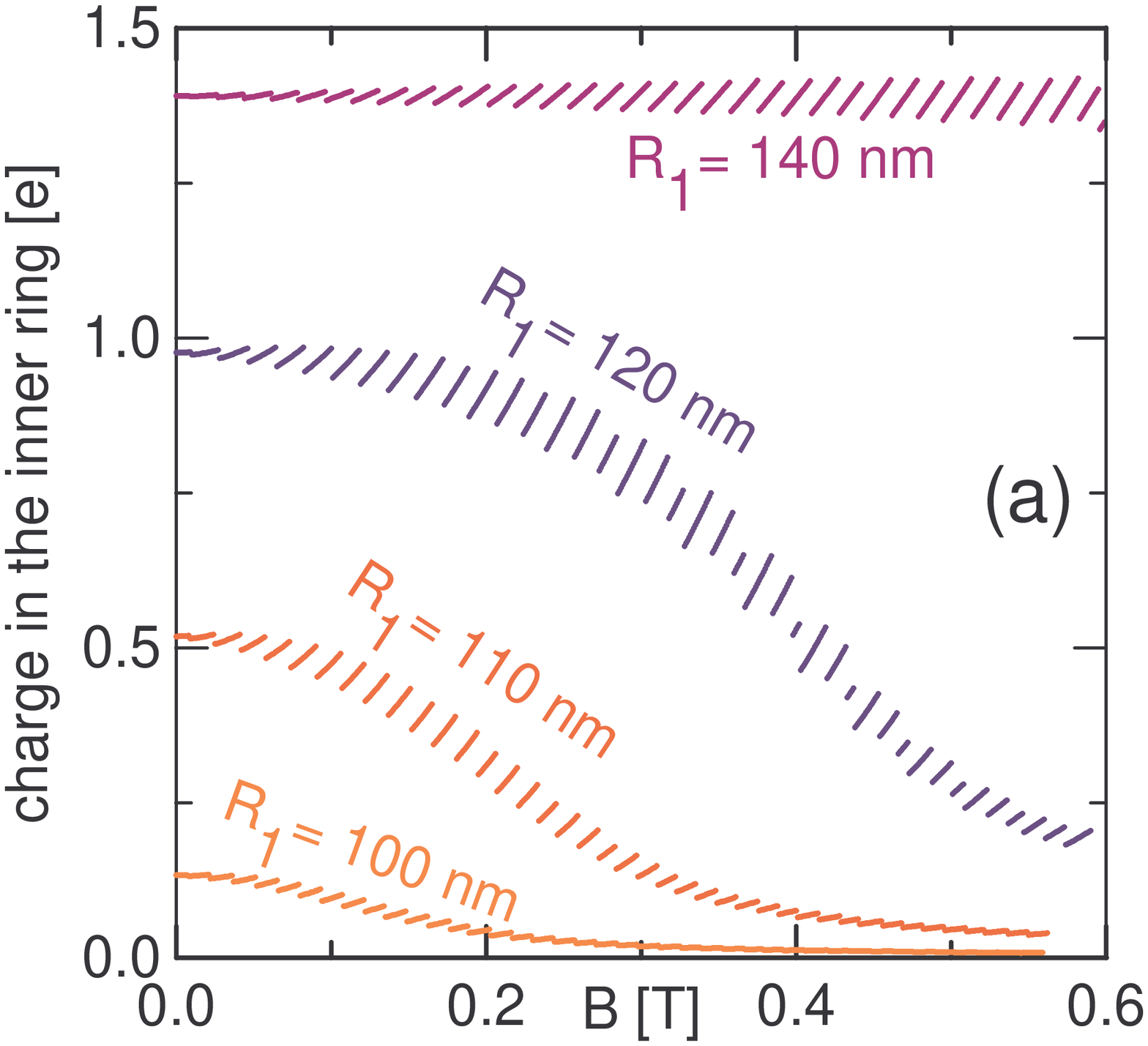}\hfill}
                \hbox{\epsfxsize=50mm
                \epsfbox[36 100 557 539] {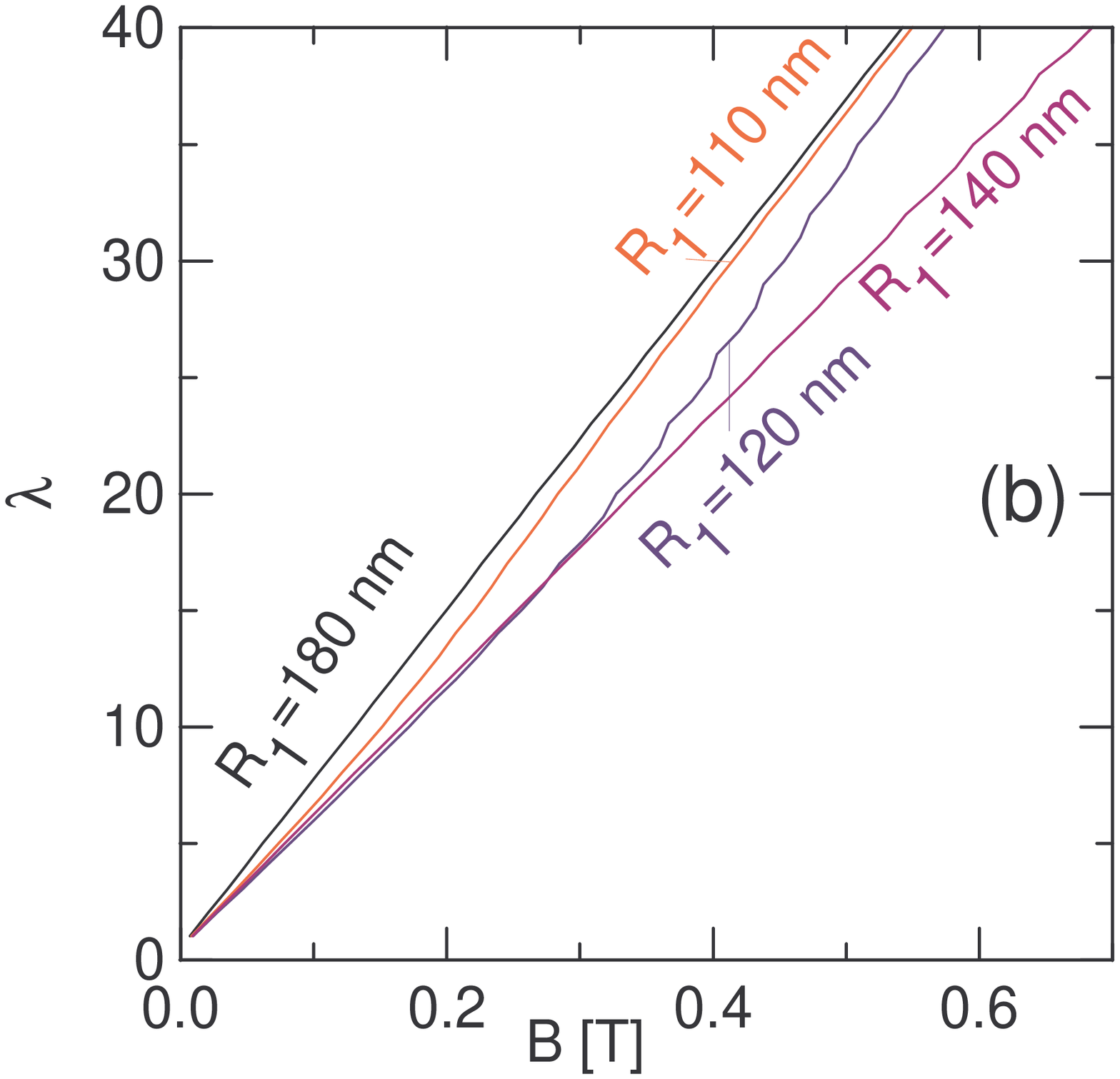}\hfill}
\caption{(color online) Three electron in two concentric rings(a)
Charge localized in the inner ring for $R_2=180$ nm and various
$R_1$ radii. (b) Upper bound for the ground-state angular
momentum.
 }
\end{figure}

\begin{figure}[htbp]
\hbox{\epsfxsize=65mm
                \epsfbox[0 0 600 800] {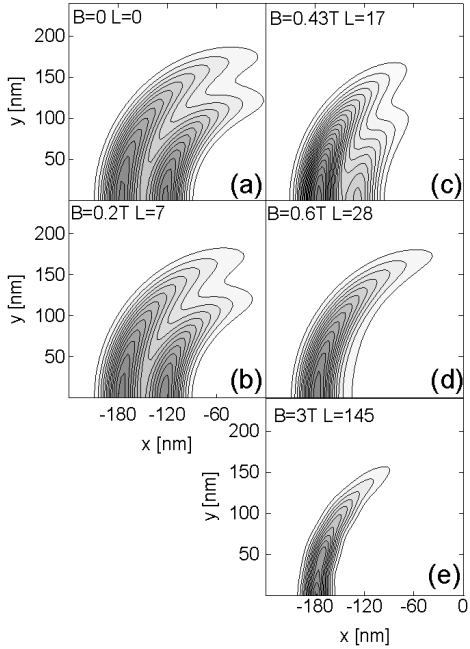}\hfill}
\caption{Pair correlation function for 2 electron ground-state in
concentric rings with radii $R_1=120$ nm and $R_2=180$ nm. One of
the electrons is fixed at the point (180 nm,0).
 }
\end{figure}

\begin{figure}[htbp]
\hbox{\epsfxsize=50mm
                \epsfbox[00 0 400 870] {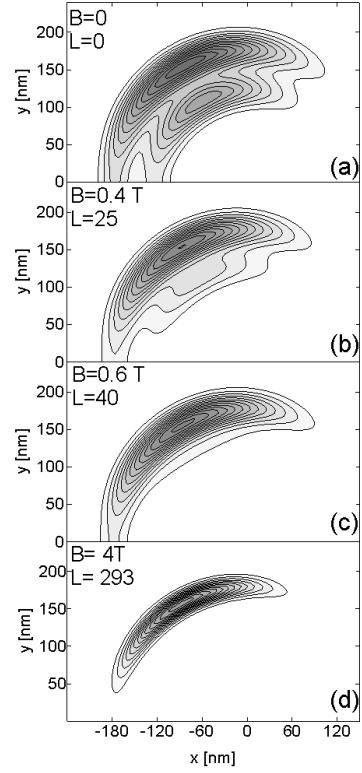}\hfill}
\caption{Same as Fig. 7, but now for 3 electrons.
 }
\end{figure}

\begin{figure*}[htbp] \hbox{\hbox{\epsfysize=64mm
                \epsfbox[27 71 559 601] {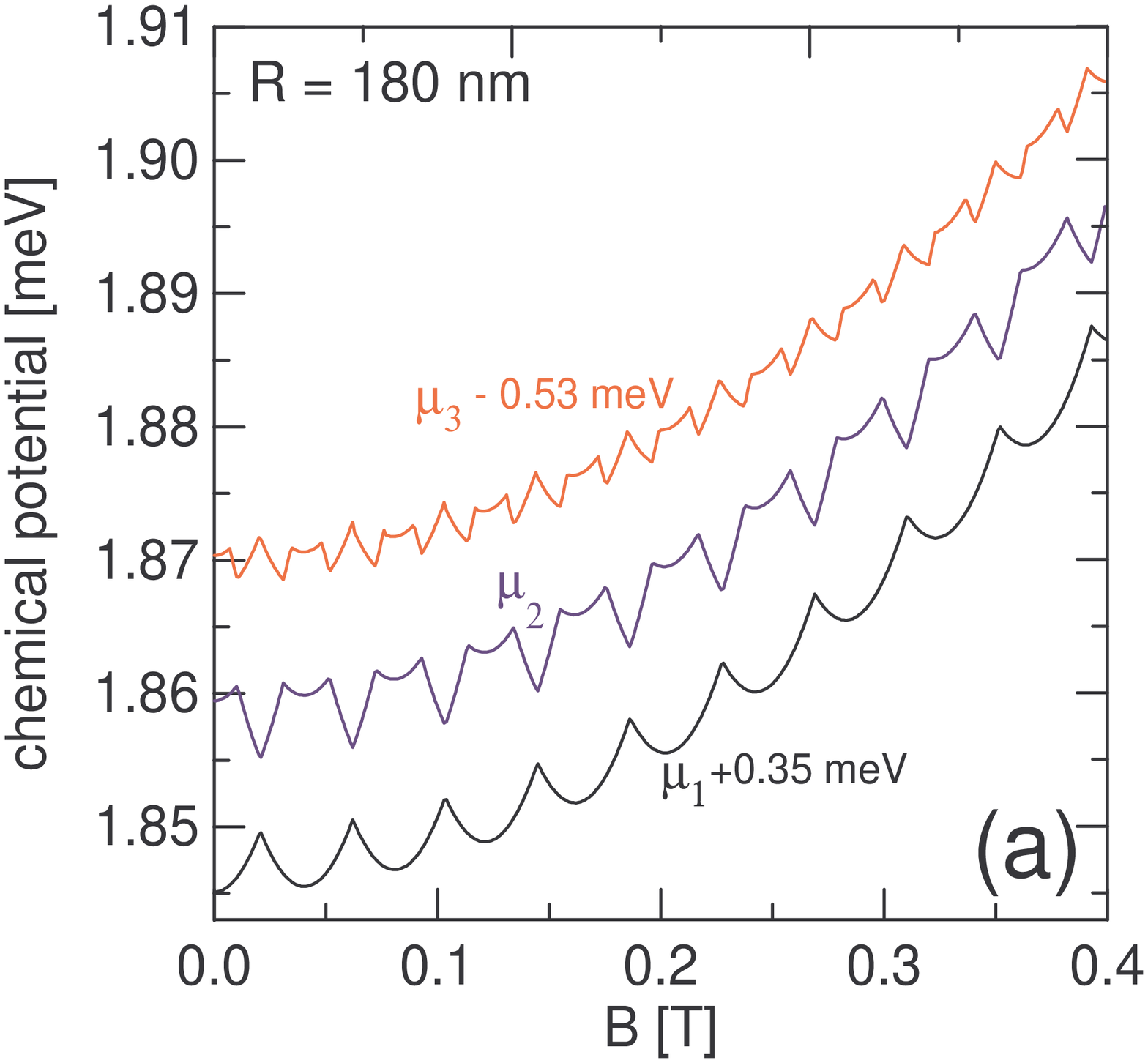}\hfill}
                \hbox{\epsfysize=64mm
                \epsfbox[20 140 551 673] {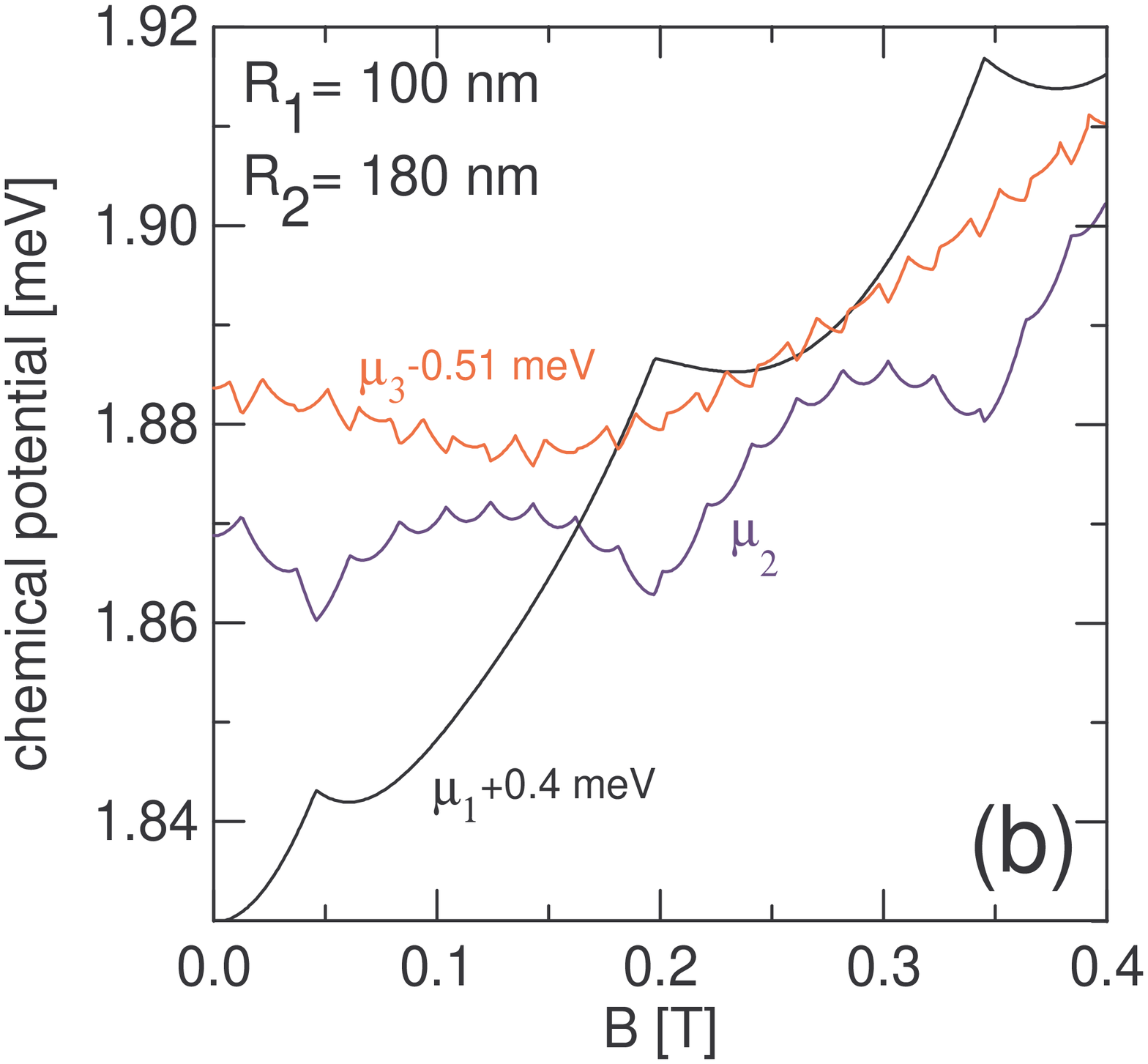}}\hfill}
                \hbox{\epsfysize=65mm                 \epsfbox[20 35 550 569] {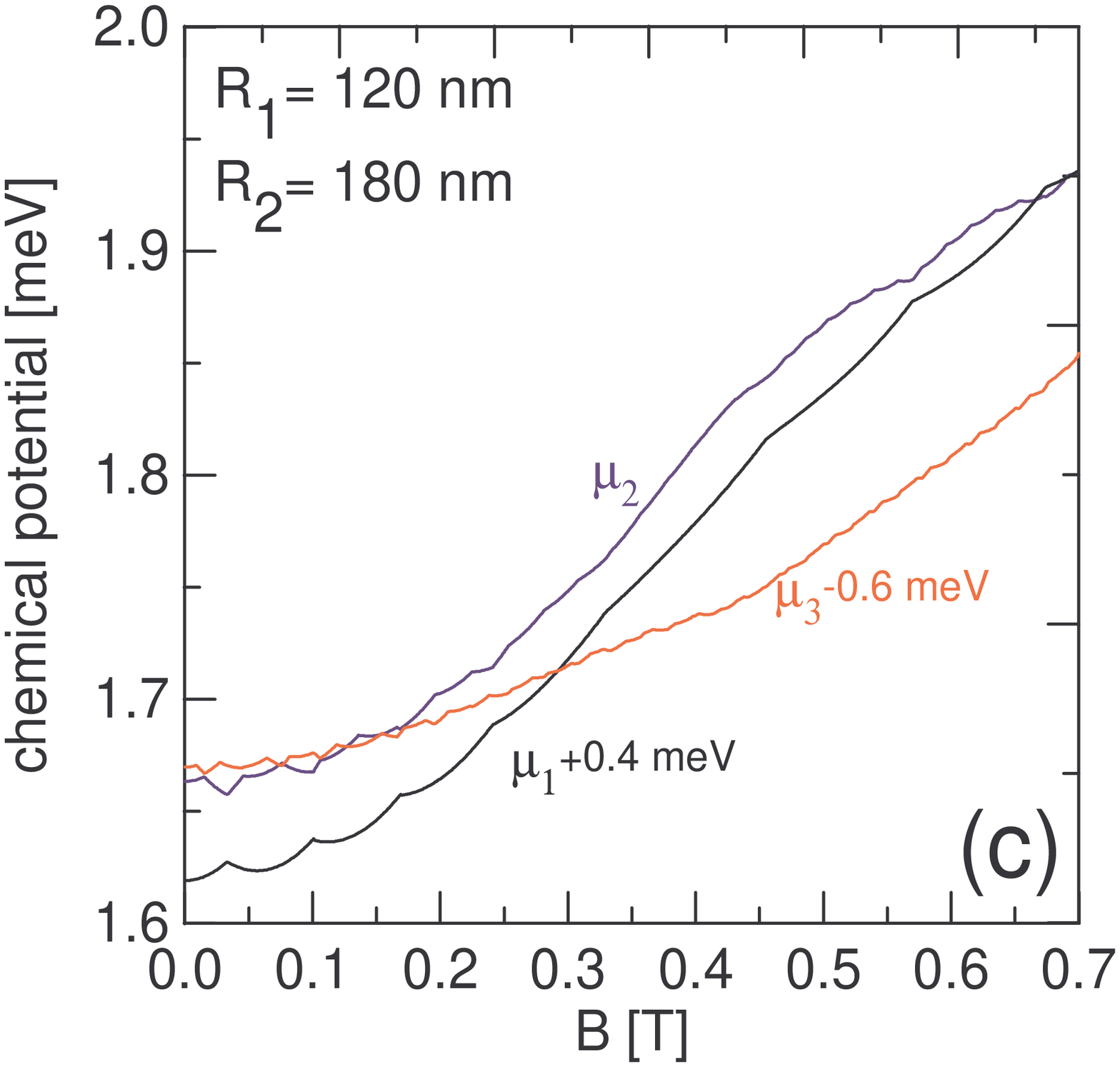}
                \epsfysize=62mm                 \epsfbox[8 310 741 779]
                {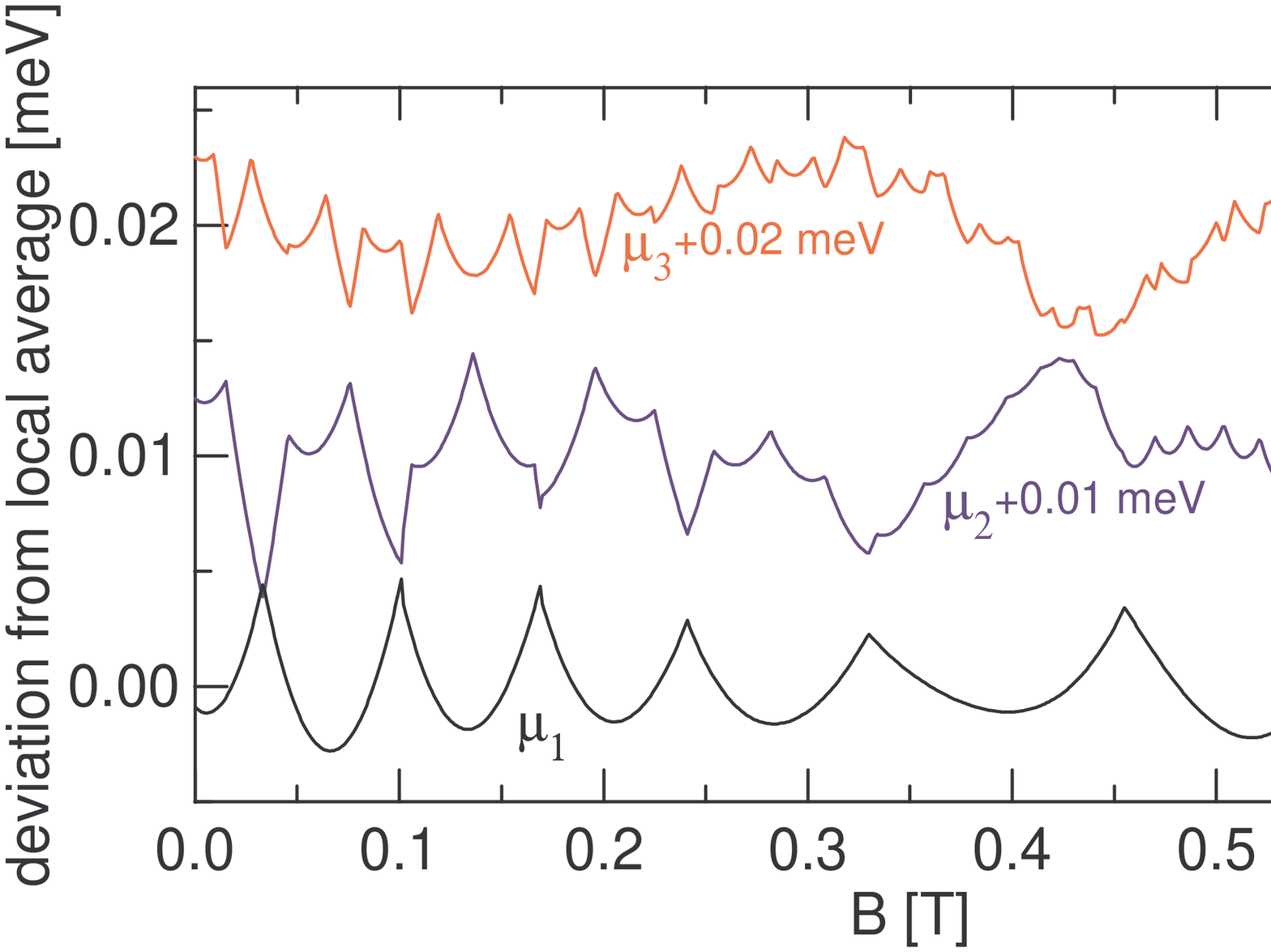}}

\caption{(color online) Magnetic field dependence of the chemical
potentials for 1, 2 and 3 electron systems in a single ring of
radius $R=180$ nm (a), in double concentric rings of external ring
radius $R_2=180$ nm and internal ring radius $R_1=100$ nm (b) and
$R_1=120$ nm. Chemical potentials for 1 and 3 electrons have been
shifted for clarity. (d) Deviation of the chemical potentials from
the local average (see text) for plot (c). }
\end{figure*}

\section{Few-electron eigenstates}
Let us now discuss the effect of the electron-electron interaction
on the ground-state properties of few-electron systems. We find
that for the interacting two-electron system the ground-state
angular momentum takes on all the subsequent integer values, like
for a single quantum ring. The upper bound for the ground state
angular momentum of the two-electron system $\lambda\ge L$ is
plotted in Fig. 5(a) as a function of the external magnetic field.
In contrast to the single-electron problem, no influence of the
inner ring on the ground-state angular momentum is observed for
$R_1<90$ nm. This indicates, that the Coulomb repulsion prevents
the electrons to occupy the inner ring if its radius is too small.
As a signature of the inter-ring coupling we see for $R_1=115$,
120 and 122 nm, that the ground-state angular momentum increases
initially more slowly than for the single $R=180$ nm ring
indicating the presence of electron charge in the internal ring.
At a certain value of magnetic field however, the lines change
their slope and tend to the values obtained for a single ring of
radius 180 nm. In the discussed range of the magnetic field the
inter-ring coupling for the internal ring radii $R_1=130$ and 140
nm is preserved.

Dotted lines in Fig. 5(a) show the $\lambda$ values for the
noninteracting electron couple for a single ring with $R=180$ nm
(black dots) and for the double ring with $R_1=120$ and $R_2=180$
nm (orange dots). For the single $R=180$ nm ring, the $\lambda$
values for the interacting and noninteracting cases run parallel
to one another. However, for $R=120$ nm the $\lambda$ values for
the noninteracting pair decreases its slope as the magnetic field
is increased, while for the interacting pair an increase of the
slope is observed instead. This is because for high magnetic
fields the interacting electrons tend to occupy the external ring
[cf. Fig. 5 (b)] to minimize their mutual repulsion in contrast to
the single-electron problem in which the diamagnetic term of the
Hamiltonian promotes the localization in the inner ring  (see Fig.
4).

The energy spectrum for $R_1=122$ nm, for which the localization
of the charge in the external ring appears in the most abrupt way
is plotted in Fig. 5(c). Below $B=0.4$ T one can observe two bands
of energy levels. In the ground state the spin singlets correspond
to even angular momenta and the spin triplets to odd angular
momenta. Opposite correspondence is found in the excited energy
band. The two bands approach each other near $B=0.5$ T, but never
cross. The relation between the ground state spin and the even/odd
parity of the angular momentum remains unchanged [cf. singlets and
triplets of $L=24$ marked in orange in the right upper part of
Fig. 5(c)].

The distribution of the charge between the rings in the
three-electron system is qualitatively similar to the two-electron
case. At zero magnetic field the electrons refuse to occupy the
inner ring if its radius is too small [see Fig. 6(a)]. Some
electron charge is present in the internal ring due to inter-ring
tunnelling, which is lifted by the application of the external
magnetic field.
 The ground-state angular momentum at high magnetic
field tends to the value obtained for a single, external ring [see
Fig. 6(b)]. For $R_1=140$ nm, in the range of the magnetic field
presented in Fig. 6, the inter-ring coupling is not broken [cf.
Figs. 5(a,b) for $R_1=130$ nm and $R_1=140$ nm]. In the high
magnetic field limit, when the magnetic length becomes small
compared to the size of the confining nanostructures, the charge
distribution in few-electron systems can be identified\cite{SPBA}
with the lowest-energy configuration of a classical
system\cite{bedanov} of point-charge particles. Therefore, one
should expect that in our model, assuming equal depths of both
rings, the few-electron system will eventually become entirely
localized in the external ring at still higher magnetic fields.

Next, we study the evolution of the ground-state electron-electron
correlations with increasing magnetic field. For this purpose we
consider the pair-correlation function plots given in Figs. 7 and
8 for two and three-electron systems, respectively. The position
of one of the electrons is fixed in the middle of the external
ring, namely in the point (180 nm,0). For two electrons at zero
magnetic field the second electron is found with an almost equal
probability in the outer and inner rings opposite to the fixed
electron [Fig. 7(a)]. For 0.6 T [Fig. 7(d)] the second electron
occupies mainly the external ring with a small leakage of the
probability density to the internal ring [cf. also the orange line
in Fig. 5(b)]. On the other hand, in the three-electron system at
$B=0$ there is already a pronounced shift of the pair-correlation
function to the external ring [Fig. 8(a)]. Figs. 7 and 8 show that
the infinite magnetic field limit is obtained in two steps: first
the charge is removed from the internal ring and then the angular
correlations between the electrons start to increase. The Wigner
type of localization, i.e., separation of electron charges in the
internal coordinates, increases with each ground-state angular
momentum transition tending to the point-charge limit.

The above discussed AB oscillations associated with the angular
momentum transitions can be measured through the magnetic field
dependence of the conductance\cite{PRLK} as performed in
phase-sensitive transport spectroscopy. Such transport
measurements require contacts to be attached to the nanostructure.
Connection of terminals to rings formed by the surface oxidation
technique\cite{PRLK,Fuhrer} is straightforward. On the other hand
attachment of electrodes to self-assembled rings\cite{Lorke,Nano}
has not been reported so far. However, the ground-state angular
momentum transitions can still be extracted from the chemical
potential as measured in a capacitance experiment.\cite{Lorke} The
magnetic field dependence of the chemical potentials $\mu_N$,
defined as the ground-state energy difference of $N$ and $N-1$
electrons, is presented in Fig. 9. Fig. 9(a) shows the chemical
potential for a single quantum ring of radius 180 nm. For a single
electron the chemical potential is equal to the ground-state
energy. The potential exhibits cusps having a "$\Lambda$" shape at
the angular momentum transitions. These $\Lambda$ cusps are
translated into "V" shaped cusps of the chemical potential for the
two-electron system. The angular momentum transitions in the
two-electron system are twice as frequent\cite{CEPL} as for $N=1$,
hence in the $\mu_2$ plot we observe two $\Lambda$ cusps per one V
cusp. Similarly, in the cusps pattern of the three-electron
chemical potential we obtain three $\Lambda$'s per two V's. Below
0.7 T for the double ring structure with $R_1=140$ and $R_2=180$
nm, we obtain qualitatively the same spectrum of a single-ring
type, only the AB oscillations period is increased due to the
reduced effective $R$ value. This is because for $R_1=140$ nm the
inter-ring coupling is not broken by the magnetic field for
$B<0.7$ T [see Figs. 4, 5(a), and 6(a)]. The occupied orbitals are
equally distributed between the rings.

Fig. 9(b), for the doubled ring with internal radius $R_1=100$ nm,
corresponds to the situation when a small magnetic field localizes
the single-electron ground states in the internal ring and ejects
the entire charge of the two- and three- electron systems to the
external ring [see Figs. 5(b) and 6(a)]. As a consequence, for
$\mu_2$ we observe seven to eight $\Lambda$ cusps between each
couple of V's. On the other hand, the pattern of cusps in the
chemical potential of the three-electron system resembles the
single-ring case [Fig. 9(a)], only below $B<0.1$ T a small
perturbation of the pattern is observed.

Fig. 9(c) shows the chemical potentials for $R_1=120$ nm, for
which the inter-ring tunnel coupling is quite significant at
$B=0$, but becomes suppressed in the studied range of magnetic
field [see Figs. 4, 5(a,b), 6] for all considered $N$. Note, that
for $N=1$ and 2 the range of the chemical potential modification
by the magnetic field is an order of magnitude larger than for a
single ring [see Fig. 9(a)]. A distinctly larger range of chemical
potential variation can also be noticed for $N=1$ in Fig. 9(b).
This increase is due to the magnetic field lifting of the
inter-ring coupling present at $B=0$. For larger $N$ the Coulomb
repulsion weakens the tunnel coupling at $B=0$, which explains the
weaker dependence of the envelopes of $\mu_3$ and $\mu_2$ in Fig.
9(b) and $\mu_3$ in Fig. 9(c).

In order to extract the fine features of the cusps pattern we
fitted slowly varying $6^{\mathrm th}$ order polynomials to the
chemical potentials in Fig. 9(c) and than subtracted from $\mu_N$
this local average provided by the fitted polynomial. The result
is displayed in Fig. 9(d). For $N=1$ we see an enlargement of the
AB oscillation period as the electron becomes localized in the
inner ring. The low magnetic field $\Lambda-V$ cusp sequences for
$N=2$ and 3 resemble the single ring localization [see Fig. 9(a)].
For $B>0.45$ T when both electrons are ejected to the external
ring and the single electron is localized in the inner ring we see
in $\mu_2$ several $\Lambda$'s per one V, like in Fig. 9(b). For
$\mu_3$ the single-ring type of pattern is found above $B>0.45$ T.
In the transition region ( 0.35 T $<B<$0.45 T)  the cusp structure
is less pronounced. This is due to the fact that in the $B$ range
corresponding to the transition of the electrons to the external
ring, the angular momentum increases very fast tending toward the
angular momentum of the ground state in the single quantum ring
[see Figs. 5(a,c) and 6] of radius $R=180$ nm.
\section{Summary and Conclusions}
We studied the coupling between concentric rings for the
few-electron eigenstates using the exact diagonalization approach.
We find that the strength of the tunnel coupling decreases with
angular momentum since the centrifugal potential favors the
localization of the electrons in the external ring. At high
magnetic field, for which the ground state corresponds to high
angular momentum, the tunnel coupling between the rings is
suppressed and the energy spectrum becomes decoupled into spectra
of separate external and internal rings. The ground state for the
single-electron becomes entirely localized in the inner ring due
to the diamagnetic term of the Hamiltonian, enhancing the
localization of the electron orbits. In contrast, the few-electron
states at high magnetic field become localized in the external
ring to minimize their mutual Coulomb repulsion. In our model,
assuming a similar radial confinement potential near the centers
of both rings, we find that the order of the spin-orbital
ground-state symmetries is not perturbed by the inter-ring
coupling, only the stability intervals of the subsequent
ground-states are affected by the coupling. The modification of
the electron distribution between the external and internal rings
is translated into the frequency of the ground-state angular
momentum transitions on the magnetic field scale. The electron
distribution can be extracted from the cusp patterns of the
single-electron charging lines, i.e., the chemical potential
dependence on the magnetic field. Suppression of the tunnel
inter-ring coupling and localization of the ground-states in one
of the rings under the influence of a magnetic field is
accompanied by a distinctly stronger increase of the chemical
potentials compared to the charging spectra in which the charge
distribution between the rings is not modified.

 {\bf Acknowledgments}
This work was supported by the Flemish Science Foundation (FWO-Vl)
and the Belgian Science Policy and the EU-network of excellence
SANDiE. B.S was supported by the EC Marie Curie IEF project
MEIF-CT-2004-500157.

\end{document}